\documentclass[aip, amsmath, amssymb, reprint, longbibliography]{revtex4-2}
\usepackage{graphicx}
\usepackage{dcolumn}
\usepackage{bm}
\usepackage{color}
\usepackage{braket}
\usepackage{amsmath,amssymb,amsfonts,bm}

\begin{document}

\title{Quasiparticle electronic structure of phthalocyanine:TMD interfaces from first-principles $GW$}
\author{Olugbenga Adeniran}
\author{Zhen-Fei Liu}
\email{zfliu@wayne.edu}
\affiliation{Department of Chemistry, Wayne State University, Detroit, Michigan 48202 USA}

\date{\today}

\begin{abstract}
Interfaces formed between monolayer transition metal dichalcogenides (TMDs) and (metallo)phthalocyanine molecules are promising in energy applications and provide a platform for studying mixed-dimensional molecule-semiconductor heterostructures in general. An accurate characterization of the frontier energy level alignment at these interfaces is key in the fundamental understanding of the charge transfer dynamics between the two photon absorbers. Here, we employ the first-principles substrate screening $GW$ approach to quantitatively characterize the quasiparticle electronic structure of a series of interfaces: metal-free phthalocyanine (H$_2$Pc) adsorbed on monolayer MX$_2$ (M=Mo, W; X=S, Se) and zinc phthalocyanine (ZnPc) adsorbed on MoX$_2$ (X=S, Se). Furthermore, we reveal the dielectric screening effect of the commonly used $\alpha$-quartz (SiO$_2$) substrate on the H$_2$Pc:MoS$_2$ interface, using the dielectric embedding $GW$ approach. Our calculations furnish the first set of $GW$ results for these interfaces, providing structure-property relationship across a series of similar systems and benchmarks for future experimental and theoretical studies.
\end{abstract}

\maketitle
\section{Introduction}
Interfaces formed between a pair of semiconductors feature intriguing electronic and optoelectronic properties due to the alignment of the energy levels of one component with respect to the other at the interface. The direction of the charge transfer depends on the type of the heterojunction \cite{2010}, which in turn depends on the electronic structure of the interface. Mixed-dimensional heterostructures formed between 0D molecules and 2D substrates \cite{Jariwala2017, Amsterdam2019}, in particular organic molecules deposited on monolayer transition metal dichalcogenides (TMDs) \cite{Huang2018, Wang2018, Sun2019}, have attracted much attention in recent years due to their promising applications in optoelectronic devices \cite{Choi2016}, photovoltaics \cite{Jariwala2016}, and photocatalysis \cite{Yu2020}. Among all molecular adsorbates, (metallo)phthalocyanines are notably interesting \cite{Ling2014,Gopakumar2004} by virtue of their structural planarity, large $\pi$-conjugation, photon-absorbing capability \cite{Mack2011}, fruitful surface chemistry \cite{Gottfried2015}, as well as great tunability in electronic and optical properties via a change of the metal center \cite{ArilloFlores2013,Liu2014,Zhou2021}. Once the two photon-absorbers - a (metallo)phthalocyanine molecule and a monolayer TMD - are placed together to form an interface, the underlying electronic structure, i.e., the relative alignment of energy levels of the two components dictates the mechanism and direction of the charge transfer across the interface, giving rise to distinct optoelectronic properties. To quantitatively characterize the electronic structure at these interfaces, various experimental efforts have been put forward \cite{Amsterdam2019,Choudhury2017,Kafle2019,Padgaonkar2019,Ahn2018}, while benchmark results and a systematic account of the trends are still missing, which constitute the main goals of this paper.

Complementary to experimental techniques, first-principles calculations play a unique role in elucidating the electronic structure and structure-property relationship, via modelling of atomistically well-defined systems. Notably, the energy levels at a heterogeneous interface pertinent to charge transfer are quasiparticle levels, whose accurate characterization in principle requires methods beyond the conventional density functional theory (DFT). Although many-body perturbation theory (MBPT), such as the $GW$ formalism \cite{Hedin1965,Strinati1982,Hybertsen1986}, has been very successful \cite{Malone2013,Kharche2014,Setten2015} in resolving the gap problem of DFT \cite{Perdew1982,Yang2012}, its relatively high computational cost hinders its routine applications in large systems such as the interfaces formed between (metallo)phthalocyanines and monolayer TMDs. As a result, most prior computational studies of the phthalocyanine:TMD interfaces still employ DFT, with different complexities such as semi-local functionals \cite{Choudhury2017,Yin2017}, hybrid functionals \cite{Liu2018, Amsterdam2019} or the DFT+$U$ approach \cite{Haldar2018}.

The key ingredient in $GW$ that is responsible for an accurate determination of interfacial energy level alignment is the so-called surface polarization \cite{Neaton2006,Thygesen2009}, or equivalently, dielectric screening due to the substrate. To effectively capture the dielectric screening while reducing the computational cost, the substrate screening $GW$ was proposed in Ref. \citenum{Liu2019}, which was shown to be accurate for weakly coupled interfaces with negligible orbital hybridization \cite{Ugeda2014, Liu2019, Xuan2019, Adeniran2020, Shunak2021}. In this work, we apply this approach to a series of interfaces: metal-free phthalocyanine (H$_2$Pc) adsorbed on monolayer MX$_2$ (M=Mo, W; X=S, Se) and zinc phthalocyanine (ZnPc) adsorbed on MoX$_2$ (X=S, Se). We focus on the quasiparticle electronic structure, especially the frontier energy level alignment between the valence band maximum (VBM) and the conduction band minimum (CBM) of the monolayer TMD with the highest occupied molecular orbital (HOMO) and lowest unoccupied molecular orbital (LUMO) of the H$_2$Pc or ZnPc. Our results reveal structure-property relationship and provide $GW$-quality benchmark results for future studies.

Furthermore, in most experimental studies, a substrate is used to support the composite molecule:TMD interface system from the bottom, with a commonly used one being $\alpha$-quartz (SiO$_2$) \cite{Choi2016,Pak2015,Ghimire2018,Ghimire2018,Ahn2018,Mutz2020}. Due to the dielectric screening effect, substrates could greatly affect the electronic properties of the adsorbate on top of it \cite{Zheng2016,Zibouche2021,Ryou2016,Qiu2017}. In this work, we use the dielectric embedding $GW$ approach \cite{Liu2020} to address the dielectric effect of the SiO$_2$ substrate on the H$_2$Pc:MoS$_2$ interface. We show that with additional screening from the SiO$_2$ underneath, the energy level alignment at the H$_2$Pc:TMD interface is modulated considerably. Our results on the embedded H$_2$Pc:MoS$_2$ system agree quantitatively with experimental measurements \cite{Mutz2020}.

This paper is organized as follows. In Sec. \ref{sectionii}, we detail the computational methodology and parameters. In Sec. \ref{sectioniii}, we present our results in two aspects: the structure-property relationship across a series of systems and the dielectric effect of the SiO$_2$ substrate on the H$_2$Pc:MoS$_2$ interface. We then conclude in Sec. \ref{sectioniv} with brief remarks. The Appendix is devoted to draw a quantitative connection between two $GW$-based methods that we use, interface $GW$ and projection $GW$, to supplement our discussion in Sec. \ref{sectioniiib}. 

\section{Methodology}
\label{sectionii}
As a first step, we relax the in-plane lattice parameter and atomic coordinates of each monolayer TMD unit cell, using the vdw-DF-cx functional \cite{Berland2014}. This is the functional that we will use to relax the structure of the interfaces, so we also employ it here to ensure consistency. The calculation uses a $\mathbf{k}$-mesh of $18 {\times} 18 {\times} 1$ and a kinetic energy cutoff of 100 Ry. All DFT relaxations employ the optimized norm-conserving Vanderbilt (ONCV) pseudopotentials \cite{Schlipf2015,Hamann2013} and the \texttt{Quantum ESPRESSO} package \cite{Giannozzi2009}. The resulting lattice constants are 3.15 \AA, 3.29 \AA, 3.15 \AA, and 3.28 \AA~for monolayer MoS$_2$, MoSe$_2$, WS$_2$, and WSe$_2$, respectively. These results agree very well with experimental measurements, which yield 3.15 \AA~\cite{Wakabayashi1975}, 3.30 \AA~\cite{James1963}, 3.15 \AA~\cite{Schutte1987}, and 3.28 \AA~\cite{Schutte1987} for the four systems, respectively.

\begin{figure}[h!]
\centering
\includegraphics[width=3.3in]{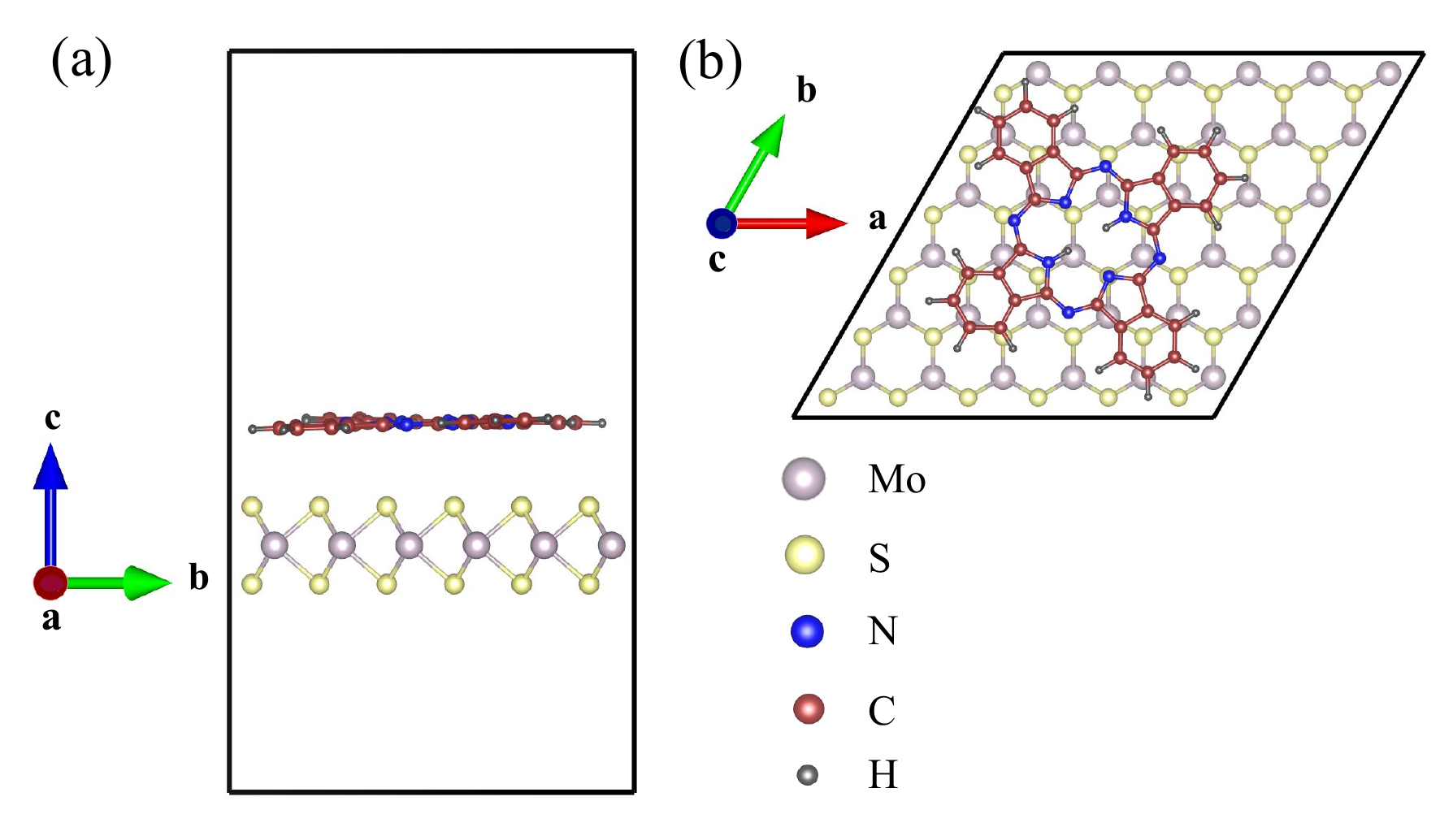}
\caption{(a) A side view and (b) a top view of the optimized H$_2$Pc:MoS$_2$ structure. The black boxes represent periodic boundary conditions. This figure is rendered using VESTA\cite{Momma2008}.}
\label{fig:figure1}
\end{figure}

After the monolayer unit cell relaxation, we build $6 {\times} 6$ supercells and place one H$_2$Pc molecule flat on each substrate and one ZnPc molecule flat on the MoS$_2$ and MoSe$_2$ substrates to form six interface systems. Each interface simulation cell is 30.0 \AA~along the $c$ direction and includes about 23 \AA~of vacuum. During the relaxation of the interface, the atoms belonging to the substrate are kept fixed in their relaxed monolayer positions, to ensure the exactness of the subsequent reciprocal-space folding of the non-interacting polarizability. The coordinates of the atoms belonging to the adsorbate molecule are full relaxed until all residual forces are below 0.05 eV/{\AA}. The relaxations are carried out using the vdw-DF-cx functional\cite{Berland2014}, a $\mathbf{k}$-mesh of $3 {\times} 3 {\times} 1$, and a kinetic energy cutoff of 70 Ry. We found an adsorption height of about 3.0 \AA~for each system we study, similar to the result of a prior calculation \cite{Amsterdam2019} (3.3 {\AA}). Fig. \ref{fig:figure1} shows the relaxed H$_2$Pc:MoS$_2$ structure in two different views.

Considering the large size of the interfaces and the associated computational cost of conventional $GW$, we apply the substrate screening $GW$ approach \cite{Liu2019} for all systems, and have explicitly benchmarked this approach against direct $GW$ calculations of the H$_2$Pc:MoS$_2$ and H$_2$Pc:MoSe$_2$ systems. All $GW$ calculations employ a mean-field starting point using the Perdew-Burke-Ernzerhof (PBE) functional \cite{Perdew1996}, the Hybertsen-Louie generalized plasmon-pole model \cite{Hybertsen1986} for the frequency dependence of the dielectric function, the semiconductor screening for the treatment of the $\mathbf{q}\to 0$ limit, the slab Coulomb truncation  \cite{IsmailBeigi2006} for the removal of spurious long-range interactions along the $c$ direction, and the static remainder \cite{Deslippe2013} in the self-energy calculation to improve convergence, as implemented in the \texttt{BerkeleyGW} package\cite{Deslippe2012}. We note that our calculations do not include the spin-orbit coupling, which is known to cause a 0.4-0.5 eV splitting for WS$_2$ and WSe$_2$ and a 0.1-0.2 eV splitting for MoS$_2$ and MoSe$_2$ in the valence band \cite{Ramasubramaniam2012}.

Here we list the computational parameters involved in the substrate screening $GW$ calculations. For the calculation of the non-interacting polarizability of the substrate unit cell, $\chi^0_{\rm sub}$, we use a $\mathbf{q}$-mesh of $18 {\times} 18 {\times} 1$, a 5 Ry dielectric cutoff, and 200 bands in the summation. For the treatment of the $\mathbf{q} \to 0$ limit, 30 bands on a shifted $\mathbf{q}$-grid are used. For the calculation of the non-interacting polarizability of the adsorbate molecule, $\chi^0_{\rm mol}$, we use a simulation cell that is the same size along $a$ and $b$ as the interface but is much smaller in size (10 \AA) along $c$. We use a $\mathbf{q}$-mesh of $3 {\times} 3 {\times} 1$, a 5 Ry dielectric cutoff, and 2400 bands in the summation. For the treatment of the $\mathbf{q} \to 0$ limit, 360 bands on a shifted $\mathbf{q}$-grid are used. After that, the $\chi^0_{\rm sub}$ is folded in the reciprocal space to a $6\times6$ supercell and the $\chi^0_{\rm mol}$ is mapped in the real space to the interface simulation cell, following Ref. \citenum{Liu2019}. These quantities are then combined at each $\mathbf{q}$-point to approximate $\chi^0_{\rm tot}$, the non-interacting polarizability of the interface. It is then inverted to generate the dielectric function in the interface simulation cell, after which the self-energies are computed following the standard procedure, where we use a $\mathbf{k}$-mesh of $3 {\times} 3 {\times} 1$, a 5 Ry dielectric cutoff, and 7200 bands of the interface in the summation for the Green's function.  

We consider two types of self-energy calculations for each interface, in line with our prior works. (i) ``Interface $GW$'' calculations, where we compute the expectation value of the self-energy operator using $\phi^{\rm tot}$, an orbital of the interface, i.e., $\braket{\phi^{\rm tot}|\Sigma|\phi^{\rm tot}}$. (ii) ``Projection $GW$'' calculations as proposed in Refs. \citenum{Tamblyn2011, Chen2017}, where we compute the expectation value of the self-energy operator using an orbital of the freestanding substrate ($\phi^{\rm sub}$) or that of the freestanding monolayer of adsorbate ($\phi^{\rm mol}$), i.e., $\braket{\phi^{\rm sub}|\Sigma|\phi^{\rm sub}}$ for the former and $\braket{\phi^{\rm mol}|\Sigma|\phi^{\rm mol}}$ for the latter. In both approaches above, the self-energy operator $\Sigma$ is calculated using $iG^{\rm tot}W^{\rm tot}$, i.e., both $G$ and $W$ are from the interface system with $W$ calculated using the substrate screening approximation. We compare the interface $GW$ and projection $GW$ results for every system, and show that they agree very well except for the LUMO of H$_2$Pc when H$_2$Pc is adsorbed on a MoS$_2$ substrate. In the Appendix, We show that this discrepancy is due to the strong orbital hybridization and, for the first time, establish a quantitative connection between the two.

\section{Results and Discussion}
\label{sectioniii}

\subsection{Convergence Study}
\label{sectioniiia}

$GW$ calculations are known to converge slowly. We describe our convergence study in this section to show that our calculations are reasonably well converged and the level alignment results are reliable.

For the monolayer MoS$_2$ unit cell, we have checked that compared to a 10 Ry dielectric cutoff and 1000 bands in the summation, our choice of parameters (5 Ry and 200 bands) lead to a convergence in the band gap within 0.02 eV, although the individual quasiparticle energies are off by about 0.3 eV. Our prediction of the monolayer MoS$_2$ gap is 2.81 eV, in good agreement with prior $GW$ calculations\cite{Shi2013} (2.80 eV). We also note that for such low-dimensional materials, the nonuniform neck subsampling method \cite{Jornada2017} (NNS) leads to faster convergence. Although we do not use it here for the interface, our calculations of the monolayer MoS$_2$ unit cell achieve reasonably good agreement (within 0.1 eV in the gap) with NNS results with a 10 Ry dielectric cutoff. 

For the interface systems, using H$_2$Pc:MoSe$_2$ as an example, we have checked that our choice of parameters (a 5 Ry dielectric cutoff and 7200 bands) lead to a convergence of the quasiparticle energies and energy level alignments at the interface within 0.05 eV, compared to using a 7.5 Ry cutoff and 9000 bands in the summation. This choice of parameter is thus adopted for all other interface systems. 

\subsection{Molecule:TMD Interfaces}
\label{sectioniiib}
\begin{figure*}[htp]
\centering
\includegraphics[width=6in]{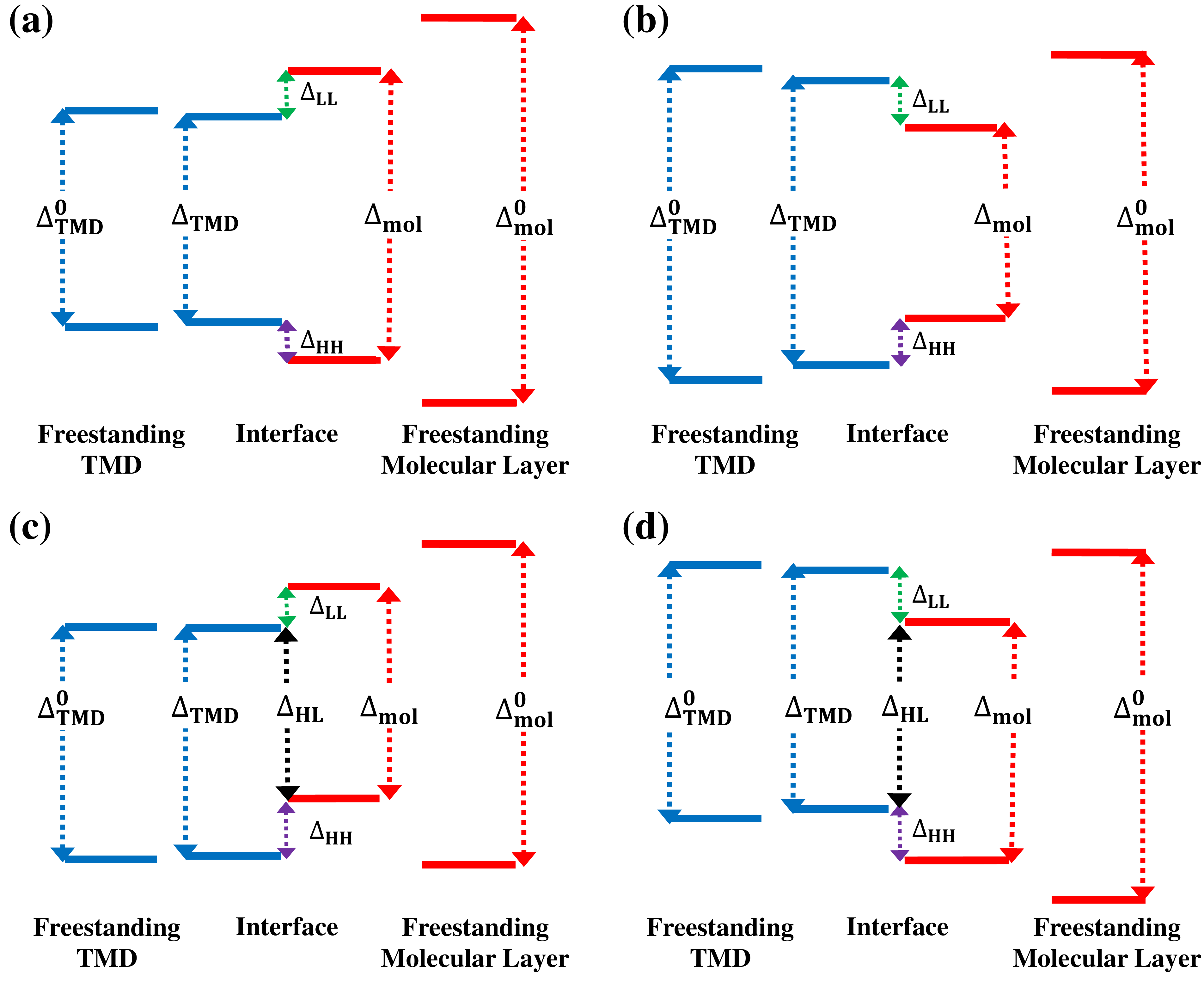}
\caption{Energy level alignment diagrams showing relevant gaps across the interface. $\Delta^0_{\rm TMD}$ ($\Delta^0_{\rm mol}$) is the band gap of the TMD monolayer (H$_2$Pc or ZnPc molecular layer) in its freestanding form, while $\Delta_{\rm TMD}$ ($\Delta_{\rm mol}$) is the gap of the TMD (H$_2$Pc or ZnPc) within the interface system. $\Delta_{\rm LL}$ ($\Delta_{\rm HH}$) is the gap between the TMD CBM (VBM) and the LUMO (HOMO) of the molecular layer. (a) and (b) display type-I heterostructures between TMD and the molecule. In (a) [(b)], both the VBM and CBM of the interface are localized on the TMD (molecule), which we denote ``type-Ia'' (``type-Ib''). (c) and (d) display type-II heterostructures between TMD and the molecule. In (c) [(d)], the VBM of the interface is localized on the molecule (TMD) and the CBM of the interface is localized on the TMD (molecule), which we denote ``type-IIa'' (``type-IIb''). In both (c) and (d),  $\Delta_{\rm HL}$ denotes the band gap of the interface.}
\label{fig:generic}
\end{figure*}

When a molecule is adsorbed on a semiconductor substrate, the interface could in principle exhibit the so-called type-I (straddling band gap), type-II (staggered band gap), or type-III (broken band gap) energy level alignment \cite{2010}. Specific to the H$_2$Pc:TMD or ZnPc:TMD interfaces studied in this work, type-I and type-II heterostructures are possible, based on our results below. Depending on the relative ordering between TMD and molecular levels, we further categorize the interfaces to type-Ia, type-Ib, type-IIa, and type-IIb, as schematically shown in Fig. \ref{fig:generic}(a)-(d), respectively. All interfaces studied in this work have direct band gap at $\Gamma$, so all interface gaps and energy level alignments are reported for the $\Gamma$ point.

\begin{table*}[htp]
\caption{Key descriptors of the electronic structure for different interface systems as calculated from DFT (using the PBE functional) and $GW$. All values are in eV. $\Delta^0_{\rm TMD}$ and $\Delta^0_{\rm mol}$ are the band gaps of the freestanding TMD monolayer and molecular layer, respectively. $\Delta^0_{\rm TMD}$ is calculated at the $K$ point for the unit cell, and $\Delta^0_{\rm mol}$ is calculated at the $\Gamma$ point. $\Delta_{\rm TMD}$, $\Delta_{\rm LL}$, $\Delta_{\rm HL}$, $\Delta_{\rm HH}$, and $\Delta_{\rm mol}$ are energy level differences within the interface systems defined in Fig. \ref{fig:generic}, all calculated at the $\Gamma$ point.}
\label{tab:Table1}
\begin{tabular}{c|c|c|c|ccccc|c}
\hline
Interface & Method & Type & $\Delta^0_{\rm TMD}$ & $\Delta_{\rm TMD}$ & $\Delta_{\rm LL}$ & $\Delta_{\rm HL}$ & $\Delta_{\rm HH}$ & $\Delta_{\rm mol}$ & $\Delta^0_{\rm mol}$ \\
 \hline
H$_2$Pc:MoS$_2$ & DFT & IIa, Fig. \ref{fig:generic}(c) & 1.79 & 1.79 & 0.09 & 1.21 & 0.58 & 1.30 & 1.39\\
& $GW$ & IIa, Fig. \ref{fig:generic}(c) & 2.81 & 2.78 & 0.14 & 2.33 & 0.45 & 2.47 & 3.86 \\
H$_2$Pc:MoS$_2$:SiO$_2$ & $GW$ & IIa, Fig. \ref{fig:generic}(c) & 2.81 & 2.07 & 0.68 & 1.62 & 0.45 & 2.30 & 3.86\\
H$_2$Pc:MoSe$_2$ & DFT & IIb, Fig. \ref{fig:generic}(d) & 1.55 & 1.55 & 0.24 & 1.31 & 0.06 & 1.37 & 1.37\\
 & $GW$ & Ia, Fig. \ref{fig:generic}(a) & 2.40 & 2.39 & 0.20 & $-$ & 0.18 & 2.77 & 3.86 \\
H$_2$Pc:WS$_2$ & DFT & Ib, Fig. \ref{fig:generic}(b) & 1.93 & 1.96 & 0.24 & $-$ & 0.36 & 1.36 & 1.36\\
 & $GW$ & IIa, Fig. \ref{fig:generic}(c) & 3.05 & 2.98 & 0.14 & 2.76 & 0.22 & 2.90 & 3.86 \\
H$_2$Pc:WSe$_2$ & DFT & IIb, Fig. \ref{fig:generic}(d) & 1.68 & 1.65 & 0.54 & 1.11 & 0.26 & 1.37 & 1.39\\
& $GW$ & Ia, Fig. \ref{fig:generic}(a) & 2.65 & 2.44 & 0.05 & $-$ & 0.35 & 2.84 & 3.86 \\
ZnPc:MoS$_2$ & DFT & IIa, Fig. \ref{fig:generic}(c) & 1.79 & 1.80 & 0.27 & 1.15 & 0.65 & 1.42 & 1.42\\
 & $GW$ & IIa, Fig. \ref{fig:generic}(c) & 2.81 & 2.78 & 0.59 & 2.29 & 0.49 & 2.88 & 3.91 \\
ZnPc:MoSe$_2$ & DFT & Ib, Fig. \ref{fig:generic}(b) & 1.55 & 1.55 & 0.16 & $-$ & 0.04 & 1.35 & 1.35\\
 & $GW$ & Ia, Fig. \ref{fig:generic}(a) & 2.40 & 2.38 & 0.28 & $-$ & 0.30 & 2.96 & 3.91\\
\hline
\end{tabular}
\end{table*}

In Fig. \ref{fig:generic}, we show frontier energy levels (bands) of the freestanding monolayer TMD and those of the freestanding molecular layer, together with their counterparts within the interface. Blue lines represent TMD levels and red lines represent molecular levels. We discuss the following quantities that characterize the electronic structure of these interfaces: $\Delta^0_{\rm TMD}$ ($\Delta^0_{\rm mol}$) is the band gap of the TMD monolayer (H$_2$Pc or ZnPc molecular layer) in its freestanding form, while $\Delta_{\rm TMD}$ ($\Delta_{\rm mol}$) is the gap of the TMD (H$_2$Pc or ZnPc) within the interface system. $\Delta_{\rm LL}$ ($\Delta_{\rm HH}$) is the gap between the TMD CBM (VBM) and the LUMO (HOMO) of the molecular layer, which is of interest for both type-I and type-II heterostructures, because the sign and magnitude of $\Delta_{\rm LL}$ ($\Delta_{\rm HH}$) dictate the direction and barrier for electron (hole) transfer across the interface, respectively. For the type-II heterostructures in Fig. \ref{fig:generic}(c)(d), we further consider $\Delta_{\rm HL}$, the fundamental (transport) gap of the entire interface, which is between the HOMO of the molecule (the VBM of the TMD) and the CBM of the TMD (the LUMO of the molecule) for type-IIa (type-IIb). $\Delta^0_{\rm TMD}$ is calculated at the $K$ point of the Brillouin zone for the TMD unit cell, and all other quantities are calculated at the $\Gamma$ point.

Table \ref{tab:Table1} shows the computed results for all quantities labelled in Fig. \ref{fig:generic}, from both DFT and substrate screening $GW$. To verify that the substrate screening approximation holds for the systems, we compare substrate screening $GW$ results with direct $GW$ calculations for two interfaces, H$_2$Pc:MoS$_2$ and H$_2$Pc:MoSe$_2$. This comparison shows that the substrate screening $GW$ is very accurate: for all quantities reported in Table \ref{tab:Table1}, substrate screening $GW$ leads to an agreement with direct $GW$ results within 0.05 eV. We also note that our $GW$ results on H$_2$Pc:MoS$_2$ and ZnPc:MoS$_2$ are in agreement with other calculations of the same systems (but with slightly different simulation cells) using range-separated hybrid functionals \cite{Zhou2021b}.

\begin{figure}[h!]
\centering
\includegraphics[width=3.3in]{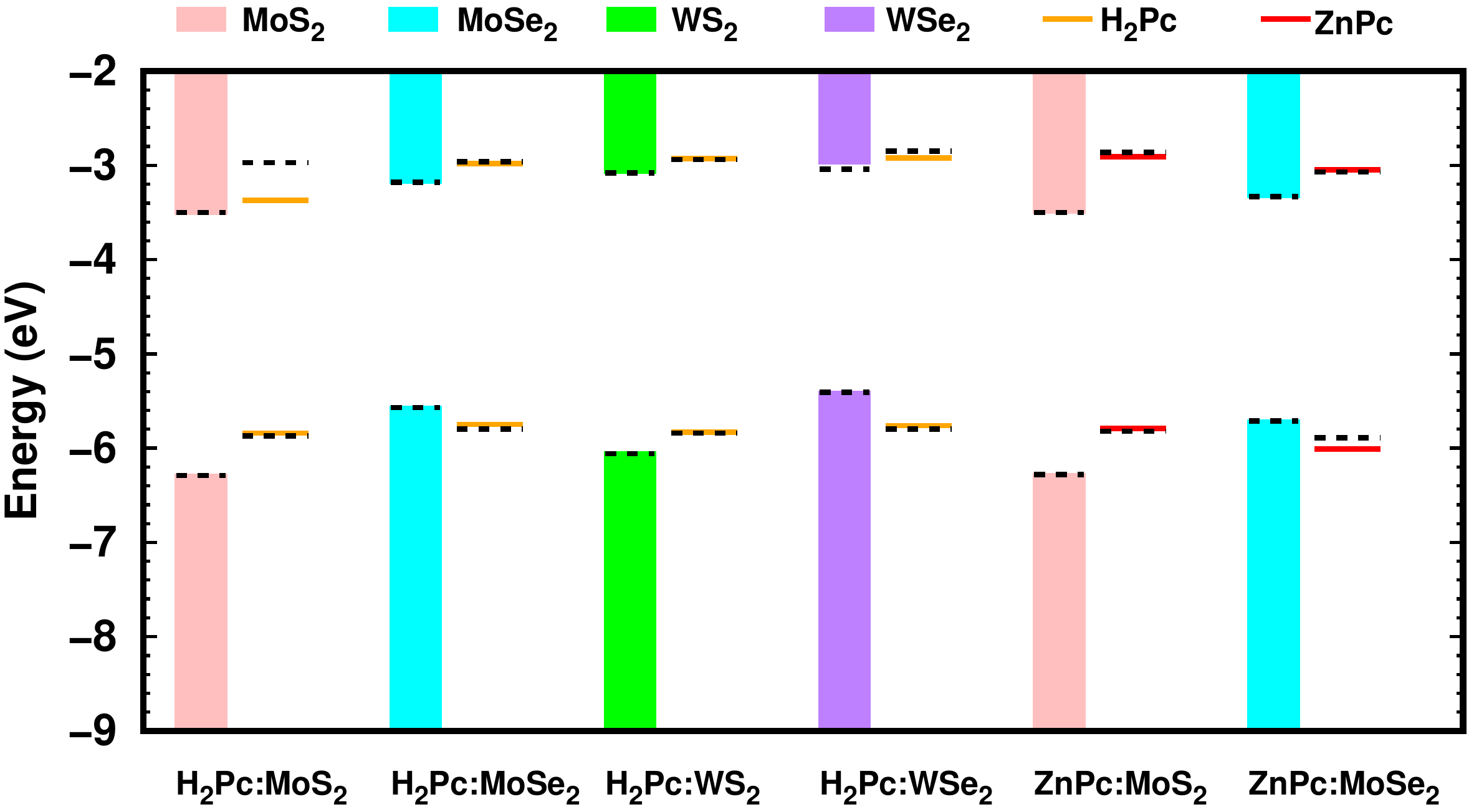}
\caption{Energy level alignment for the molecule:TMD interfaces as calculated from substrate screening $GW$. The bars denote the bands of MoS$_2$ (pink), MoSe$_2$ (blue), WS$_2$ (green), and WSe$_2$ (purple). The lines denote the energy levels of H$_2$Pc (orange) and ZnPc (red). Solid bars or lines indicate results from interface $GW$ calculations (same as those in Table \ref{tab:Table1}) and dashed lines indicate results from projection $GW$ calculations. All energy levels are measured with respect to vacuum.}
\label{fig:ela}
\end{figure}

Figure \ref{fig:ela} shows the $GW$ interfacial energy level alignment for the six heterostructures, where we use different colors to represent different substrates or molecules and all energy levels are measured with respect to a common vacuum. In Figure \ref{fig:ela}, solid bars or lines are interface $GW$ results (same as those reported in Table \ref{tab:Table1}), i.e., $\braket{\phi^{\rm tot}|\Sigma[G^{\rm tot}W^{\rm tot}]|\phi^{\rm tot}}$, where the $\phi^{\rm tot}$ is chosen as the interface orbital that mostly resembles the orbital of interest (HOMO, LUMO, VBM, or CBM) of the freestanding monolayer TMD or molecular layer, as quantitatively determined from orbital projections. Dashed lines are projection $GW$ results, i.e., $\braket{\phi^{\rm sub}|\Sigma[G^{\rm tot}W^{\rm tot}]|\phi^{\rm sub}}$ for the TMD, where $\phi^{\rm sub}$ is the VBM or CBM of the freestanding TMD, and $\braket{\phi^{\rm mol}|\Sigma[G^{\rm tot}W^{\rm tot}]|\phi^{\rm mol}}$ for the molecule, where $\phi^{\rm mol}$ is the HOMO or LUMO of the freestanding molecular layer. For all cases except the H$_2$Pc LUMO on MoS$_2$ substrate, the interface $GW$ results agree very well with projection $GW$ results, which indicates negligible orbital hybridization upon formation of the interface, such that $\braket{\phi^{\rm tot}|\phi^{\rm sub}}$ and $\braket{\phi^{\rm tot}|\phi^{\rm mol}}$ are close to unity. The special case of H$_2$Pc LUMO on MoS$_2$ substrate is discussed in the Appendix.

Below in this section, we discuss the results from three aspects: (i) the renormalization of gaps upon the formation of the interface; (ii) the qualitative difference between DFT and $GW$ in the prediction of the type of some heterostructures; and (iii) the structure-property relationship across the six different systems.

For the freestanding monolayer TMD, our $GW$ calculations yield band gaps of 2.81 eV, 2.40 eV, 3.05 eV, and 2.65 eV for MoS$_2$, MoSe$_2$, WS$_2$, and WSe$_2$, respectively. Our results are in good agreement with Ref. \citenum{Shi2013}, where sc$GW_0$ calculations of monolayer TMDs with similar lattice parameters (within 0.01 \AA~of our relaxed values) were performed, resulting in band gaps of 2.80 eV, 2.40 eV, 3.11 eV, and 2.68 eV, respectively, for the four materials. For the freestanding H$_2$Pc (ZnPc) molecular layer, our $GW$ calculations yield a HOMO-LUMO gap of 3.86 eV (3.91 eV) and an ionization potential of 6.24 eV (6.16 eV), on par with Ref. \citenum{Blase2011}, where $GW$ calculation of an isolated H$_2$Pc molecule yielded a band gap of 3.67 eV and an ionization potential of 6.08 eV. 

At the interface, the gap renormalization for the H$_2$Pc or ZnPc molecule is significant, with about 1 eV decrease in the HOMO-LUMO gap when the molecule is brought in contact with the substrate. This is consistent with well-established understanding of the surface renormalization at molecule-substrate interfaces \cite{Neaton2006,Thygesen2009}. On the contrary, Table \ref{tab:Table1} shows that TMD band gaps are still very similar to those of the freestanding phase, indicating negligible gap renormalization for the substrate, which we attribute to the relatively low coverage of the molecule on the substrate (see Fig. \ref{fig:figure1}). This situation is different from a previous semiconductor-semiconductor interface that we studied before \cite{Adeniran2020}, where we observed gap renormalization on both sides of an organic bulk heterojunction. For the H$_2$Pc:MoS$_2$:SiO$_2$ interface, the MoS$_2$ gap is significantly renormalized due to the dielectric screening effect of the extensive SiO$_2$ substrate underneath, as we will elaborate in Sec. \ref{sectioniiic} below. Needless to say, DFT clearly underestimates the relevant gaps, and does not capture the gap renormalization at the interface.

A remarkable observation here is that for some systems, DFT and $GW$ yield qualitatively different heterostructure type, as listed in Table \ref{tab:Table1}. To be specific, for H$_2$Pc:MoSe$_2$ and H$_2$Pc:WSe$_2$, DFT predicts a type-II while $GW$ predicts a type-I interface; for H$_2$Pc:WS$_2$, DFT predicts a type-I while $GW$ predicts a type-II interface; for ZnPc:MoSe$_2$, although both DFT and $GW$ predict type-I, the relative ordering of TMD and molecular levels are opposite (type-Ia and type-Ib as shown in Fig. \ref{fig:generic}). In all cases, it is the self-energy correction to the LUMO of the molecular adsorbate that pushes this orbital upward in energy, causing a change in the heterostructure type (there is an additional effect in ZnPc:MoSe$_2$, where the self-energy correction to the HOMO of ZnPc shifts this orbital downward). We note that experimental characterizations of these quasiparticle energy level alignments require techniques such as (inverse) photoemission spectroscopy, while most existing experiments \cite{Choi2016,Zhang2017,Nguyen2016,Ahn2018,Kafle2019,Amsterdam2019} focused on the description of excitons at such interfaces using, e.g., photoluminescence. Although we have not found direct experimental verification of most of the quasiparticle energy level alignments that we have computed (except for H$_2$Pc:MoS$_2$:SiO$_2$), we believe our work provides a reference point for future experiments.

Table \ref{tab:Table1} and Fig. \ref{fig:ela} reveal structure-property relationship that is helpful in materials design of similar systems. All sulfur-based interfaces form type-IIa heterostructures with HOMO (LUMO) of the adsorbed molecule lying higher than VBM (CBM) of the TMD substrate. All selenium-based ones form type-Ia heterostructures with HOMO (LUMO) of the adsorbed molecule lying lower (higher) than the VBM (CBM) of the TMD substrate. Comparing ZnPc with H$_2$Pc on the same substrates, we find that the gap renormalization of ZnPc is smaller, resulting in larger $\Delta_{\rm LL}$ and $\Delta_{\rm HH}$ values than H$_2$Pc-based interfaces. Since these values represent the charge transfer barrier across the interface, this trend suggests that the electron or hole transfer rates across H$_2$Pc-based interfaces might be generally higher than those across ZnPc-based interfaces.

\subsection{The Effect of SiO$_2$ substrate on the H$_2$Pc:MoS$_2$ Interface}
\label{sectioniiic}

In typical experimental studies, the molecule:TMD interfaces are further supported by a substrate underneath the monolayer TMD \cite{Zheng2016,Mutz2020}. It is well known that the band gaps of TMDs are sensitive to the dielectric environment provided by other adjacent 2D materials or substrates \cite{Ryou2016,Raja2017,Zibouche2021}. Therefore, it is imperative to include the dielectric screening effects from any additional substrates to achieve quantitative agreement with experimental characterization of the electronic structure of the molecule:TMD interfaces. In this work, we employ the dielectric embedding $GW$ approach as developed in Ref. \citenum{Liu2020} to include the effect of the substrate, using the H$_2$Pc:MoS$_2$ system as an example. We focus on this system because an experimental measurement is available \cite{Mutz2020} for a direct comparison.

A commonly used substrate in experimental studies of molecule:TMD interfaces is the $\alpha$-quartz (SiO$_2$) \cite{Choi2016,Pak2015,Ghimire2018,Ghimire2018,Ahn2018,Amsterdam2019,Mutz2020}. To model the composite H$_2$Pc:MoS$_2$:SiO$_2$ system, we have applied a 5\% compressive strain to the experimental lattice constant of SiO$_2$ to enforce a commensurate simulation cell with H$_2$Pc:MoS$_2$. Under this strain, the PBE band gap of 5.93 eV for bulk SiO$_2$ \cite{Tran2017} increases by about 4\% while other qualitative features of the band structure remain intact. Using the dielectric embedding $GW$ approach \cite{Liu2020}, we first compute the non-interacting polarizability of the SiO$_2$ substrate in its unit cell, then fold this quantity in reciprocal space to the supercell, and lastly, combine the folded quantity with the non-interacting polarizability of the H$_2$Pc:MoS$_2$ (which in turn is calculated using the substrate screening $GW$ \cite{Liu2019} approach). As a result, the self-energy calculation is only explicitly performed for the H$_2$Pc:MoS$_2$ interface, where the $W$ includes the dielectric effect of the SiO$_2$ substrate. This approach adds little extra computational cost compared to the $GW$ calculation of H$_2$Pc:MoS$_2$ without the SiO$_2$ substrate.

\begin{figure}[h!]
\centering
\includegraphics[width=3.1in]{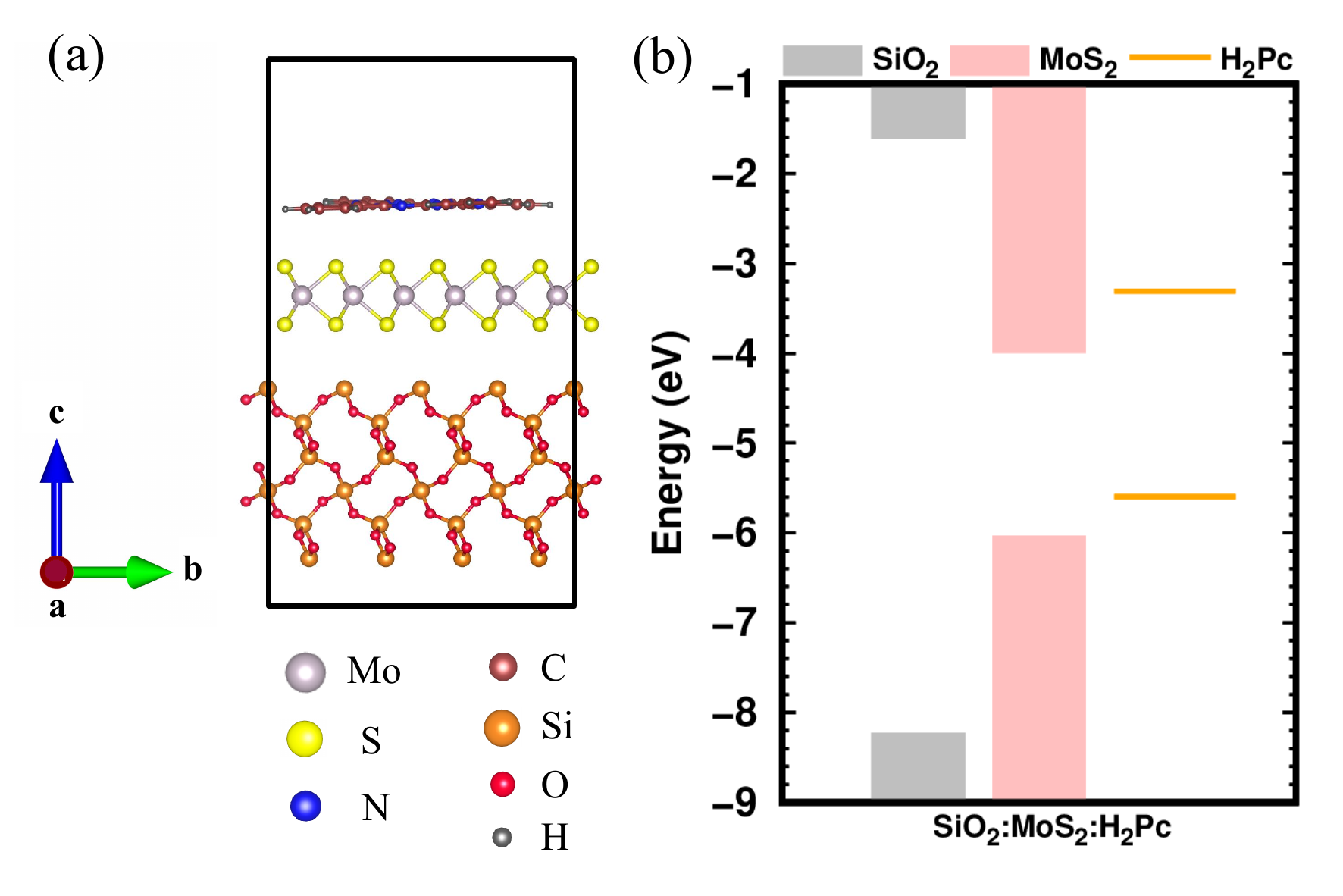}
\caption{(a) Optimized structure of the H$_2$Pc:MoS$_2$:SiO$_2$ interface. (b) Quasiparticle energy level alignment from an embedding $GW$ calculation, where the H$_2$Pc:MoS$_2$ interface is embedded into a dielectric environment of a SiO$_2$ substrate. All energy levels are measured with respect to vacuum.}
\label{fig:sio2-ela}
\end{figure}

Fig. \ref{fig:sio2-ela}(a) shows the optimized structure of the H$_2$Pc:MoS$_2$:SiO$_2$ system, where we consider a two-layer SiO$_2$ substrate with silicon termination. The separation between the SiO$_2$ and MoS$_2$ is optimized using vdw-DF-cx to be 3.5 \AA. Fig. \ref{fig:sio2-ela}(b) shows the quasiparticle electronic structure of the H$_2$Pc:MoS$_2$ interface embedded in the dielectric environment of SiO$_2$, computed using the dielectric embedding $GW$ approach \cite{Liu2020}, with key energy gaps listed in Table \ref{tab:Table1}. Compared to the H$_2$Pc:MoS$_2$ system without the SiO$_2$ substrate, the MoS$_2$ band gap is further renormalized to 2.07 eV due to the additional dielectric screening from SiO$_2$, which is in very good agreement with experiment (2.10 eV as in Ref. \citenum{Mutz2020}). The H$_2$Pc gap is further renormalized only moderately, possibly due to its large distance to the SiO$_2$ substrate. Furthermore, the $\Delta_{\rm LL}$ is considerably changed to 0.68 eV (compared to 0.14 eV without the SiO$_2$ substrate), and the $\Delta_{\rm HL}$ is changed to 1.62 eV (compared to 2.33 eV without the SiO$_2$ substrate). Interestingly, the $\Delta_{\rm HH}$ does not change compared to the case without the SiO$_2$ substrate. 

As a direct comparison with experiment, Ref. \citenum{Mutz2020} characterized the energy level alignment of the H$_2$Pc:MoS$_2$ interface deposited on SiO$_2$ substrate. Ultraviolet photoemission spectroscopy and inverse photoemission spectroscopy measurements were performed for pristine monolayer MoS$_2$ and the H$_2$Pc:MoS$_2$ interface, both on the SiO$_2$ substrate. By aligning the Fermi level of the two systems, Ref. \citenum{Mutz2020} deduced that $\Delta_{\rm TMD}=2.1$ eV, $\Delta_{\rm LL}=1.0$ eV, $\Delta_{\rm HL}=1.2$ eV, $\Delta_{\rm HH}=0.9$ eV, and $\Delta_{\rm mol}=2.2$ eV. We note that our results agree quantitatively with Ref. \citenum{Mutz2020} in $\Delta_{\rm TMD}$ and $\Delta_{\rm mol}$ values, but our computed H$_2$Pc HOMO and LUMO levels are both lower by 0.3-0.4 eV than those reported in Ref. \citenum{Mutz2020}, resulting in lower $\Delta_{\rm LL}$, lower $\Delta_{\rm HH}$ and higher $\Delta_{\rm HL}$. We discuss two possible sources for the discrepancy. First, Ref. \citenum{Mutz2020} aligned the Fermi level of the pristine MoS$_2$ (without the H$_2$Pc adsorbate) and that of the H$_2$Pc:MoS$_2$ interface. The precise position of the Fermi level within the band gap might depend on the specific experimental condition, while it is not well-defined in first-principles calculations at zero temperature. This difference between an experiment and a computation affects how the MoS$_2$ and H$_2$Pc levels are relatively aligned without affecting the band gap of each component. Second, the coverage of the H$_2$Pc adsorbate on MoS$_2$ substrate might be different between Ref. \citenum{Mutz2020} and our modelling, which might lead to different interface dipoles and consequently different $\Delta_{\rm LL}$, $\Delta_{\rm HL}$, and $\Delta_{\rm HH}$ values.

The result is a clear indication of the dielectric screening effect of the substrate on molecule:TMD interfaces. Our dielectric embedding $GW$ approach provides an accurate account of this effect without further increasing the computational cost, compared to calculations that only involve molecule:TMD interfaces. Lastly, we note in passing that our calculated $GW$ band gap is 8.69 eV for bulk SiO$_2$, in agreement with Ref. \citenum{Chen2013}, which reports a $GW$ gap of 8.77 eV. Our $GW$ gap is 7.10 eV for the freestanding bilayer SiO$_2$ with Si termination, and 6.64 eV when the bilayer SiO$_2$ is in contact with monolayer MoS$_2$, as shown in Fig. \ref{fig:sio2-ela}(b). These gaps are large enough such that the precise positioning of SiO$_2$ band edges relative to TMD bands will unlikely have large effects on the energy level alignment at the H$_2$Pc:MoS$_2$ interface. 


\section{Conclusion}
\label{sectioniv}
In this work, we have systematically and quantitatively characterized the quasiparticle electronic structure of a series of interfaces formed between (metallo)pathalocyanine molecules and monolayer TMDs, using first-principles $GW$ calculations. Besides the well-known gap renormalization of the adsorbate, we have found that in certain cases, $GW$ and DFT yield qualitatively different heterostructure types, which is of interest for future experimental validation. Furthermore, we have elucidated the dielectric screening effect of the SiO$_2$ substrate on the electronic structure of H$_2$Pc:MoS$_2$, leading to quantitative agreement with existing experiments. Our findings provide useful structure-property relationship for materials design, insight into charge transfer processes across such molecule:TMD interfaces, as well as benchmark data for future theoretical and experimental investigations.

\begin{acknowledgements}
Z.-F.L. acknowledges support from an NSF CAREER Award, DMR-2044552. O.A. acknowledges A. Paul and Carole C. Schaap Endowed Distinguished Graduate Award and Rumble Fellowship from Wayne State University. This research used computational resources at Wayne State Grid and additionally via a user project at Center for Nanoscale Materials at Argonne National Laboratory, an Office of Science user facility, which was supported by the U.S. Department of Energy, Office of Science, Office of Basic Energy Sciences, under Contract No. DE-AC02-06CH11357. Furthermore, large-scale calculations used computational resources from the National Energy Research Scientific Computing Center (NERSC), a U.S. Department of Energy Office of Science User Facility operated under Contract No. DE-AC02-05CH11231.
\end{acknowledgements}

\section*{Author Declarations}
The authors have no conflicts to disclose.

\section*{Data Availability Statement}

The data that support the findings of this study are available from the corresponding author upon reasonable request.

\appendix
\section{Deriving projection $GW$ results from interface $GW$ calculations}

We investigate the quantitative difference between projection $GW$ calculations (dashed lines in Fig. \ref{fig:ela}) and interface $GW$ calculations (solid lines in Fig. \ref{fig:ela}). They agree very well except for the H$_2$Pc LUMO on MoS$_2$ substrate, which involves strong orbital hybridization. Here, we quantitatively connect the projection $GW$ results with interface $GW$ results.

To make the discussion self-contained in the Appendix, we repeat here that in ``projection $GW$'' \cite{Tamblyn2011, Chen2017}, we compute $\braket{\phi^{\rm sub}|\Sigma|\phi^{\rm sub}}$ for the TMD substrate and $\braket{\phi^{\rm mol}|\Sigma|\phi^{\rm mol}}$ for the adsorbed molecule, where $\phi^{\rm sub}$ ($\phi^{\rm mol}$) is an orbital of the \emph{freestanding} substrate (molecule). In ``interface $GW$'', we compute $\braket{\phi^{\rm tot}|\Sigma|\phi^{\rm tot}}$ with $\phi^{\rm tot}$, where $\phi^{\rm tot}$ is an orbital of the interface that most resembles a substrate or molecular orbital (i.e., a resonance). In both cases, $\Sigma$ is the same operator that involves the $G$ and $W$ of the entire interface.

We focus on the H$_2$Pc LUMO on MoS$_2$ substrate, and expand the H$_2$Pc LUMO (calculated for the freestanding H$_2$Pc molecular layer in the same simulation cell as the interface) in terms of the orbitals of the H$_2$Pc:MoS$_2$ interface:

\begin{equation}
\ket{\phi_{\rm LUMO}^{\rm mol}} = C_1\ket{\phi_{\rm CBM+2}^{\rm tot}} + C_2\ket{\phi_{\rm CBM+8}^{\rm tot}} + C_3\ket{\phi_{\rm CBM+9}^{\rm tot}}.
\label{eq:exp1}
\end{equation}

Here, we have neglected expansion coefficients whose magnitude is below 0.05. The projection $GW$ computes $\braket{\phi^{\rm mol}_{\rm LUMO}|\Sigma|\phi^{\rm mol}_{\rm LUMO}}$. Substituting Eq. \eqref{eq:exp1}, this quantity is related to diagonal and off-diagonal matrix elements of $\Sigma$ involving the three interface orbitals on the right-hand side of Eq. \eqref{eq:exp1}. These matrix elements are from \emph{interface} $GW$ calculations, which are then connected to projection $GW$ results that involve the left-hand side of Eq. \eqref{eq:exp1}. For diagonal elements, we use the difference between the quasiparticle energies and the corresponding Kohn-Sham eigenvalues. For off-diagonal elements, we use the matrix elements of $\Sigma-V_{\rm xc}$ with $V_{\rm xc}$ being the exchange-correlation operator.

The expansion coefficients and matrix elements (in eV) are:

$C_1 = 0.6358-0.0647i, C_2 = -0.6046+0.0615i, C_3 = -0.4379+0.0446i$; 

$\braket{\phi_{\rm CBM+2}^{\rm tot}|\Sigma|\phi_{\rm CBM+2}^{\rm tot}} = 0.667$, $\braket{\phi_{\rm CBM+8}^{\rm tot}|\Sigma|\phi_{\rm CBM+8}^{\rm tot}} = 0.871$, $\braket{\phi_{\rm CBM+9}^{\rm tot}|\Sigma|\phi_{\rm CBM+9}^{\rm tot}} = 0.876$;

$\braket{\phi_{\rm CBM+2}^{\rm tot}|\Sigma|\phi_{\rm CBM+8}^{\rm tot}} = -0.227+0.222i$,

$\braket{\phi_{\rm CBM+2}^{\rm tot}|\Sigma|\phi_{\rm CBM+9}^{\rm tot}} = -0.225+0.244i$,

$\braket{\phi_{\rm CBM+8}^{\rm tot}|\Sigma|\phi_{\rm CBM+9}^{\rm tot}} = -0.104+0.001i$.

Using these values from an interface $GW$ calculation, one can calculate from Eq. \eqref{eq:exp1} that $\braket{\phi^{\rm mol}_{\rm LUMO}|\Sigma|\phi^{\rm mol}_{\rm LUMO}}$ = 1.011 eV. This is in very good agreement with the projection $GW$ calculation of the H$_2$Pc LUMO on MoS$_2$, whose self-energy correction is 1.023 eV. The above analysis successfully explains the difference between the solid and dashed lines in Fig. \ref{fig:ela}: the solid line computes $\braket{\phi_{\rm CBM+2}^{\rm tot}|\Sigma|\phi_{\rm CBM+2}^{\rm tot}}$, which only involves the first term on the right-hand side of Eq. \eqref{eq:exp1}, while the dashed line computes $\braket{\phi^{\rm mol}_{\rm LUMO}|\Sigma|\phi^{\rm mol}_{\rm LUMO}}$, which involves the left-hand side of Eq. \eqref{eq:exp1}.

As a limiting case, if only one of the coefficients in Eq. \eqref{eq:exp1} is unity, then the isolated molecular orbital and its resonance at the interface are the same (the weak-coupling limit). The projection $GW$ and the interface $GW$ will then yield identical values in the quasiparticle energy. This is, in fact, a very good approximation for most interfaces. Specific to the systems studied in this work, the two approaches agree well for every resonance except for the H$_2$Pc LUMO on the MoS$_2$ substrate.

\section{Deriving interface $GW$ results from projection $GW$ calculations}

In a similar manner to the above analysis, we can derive interface $GW$ results from projection $GW$ calculations of the substrate and of the adsorbate, which can also explain the difference between the solid and dashed lines in Fig. \ref{fig:ela}. For the H$_2$Pc LUMO on MoS$_2$ substrate, its resonance in the interface is the CBM+2 of the interface. We then consider the following expansion:

\begin{equation}
\ket{\phi_{\rm CBM+2}^{\rm tot}} = B_1\ket{\phi_{\rm LUMO}^{\rm mol}} + B_2\ket{\phi_{\rm CBM+6}^{\rm TMD}} + B_3\ket{\phi_{\rm CBM+3}^{\rm TMD}}
\label{eq:exp2}
\end{equation}

We again neglect expansion coefficients whose magnitude is below 0.05. The left-hand side of Eq. \eqref{eq:exp2} is the orbital used in an interface $GW$ calculation, and the right-hand side consists of orbitals used in projection $GW$ calculations of the substrate and of the adsorbate, respectively.

The expansion coefficients and the matrix elements (in eV) are:

$B_1 = 0.6358+0.0647i, B_2 = -0.6156-0.1232i, B_3 =-0.3509-0.0703i$;

$\braket{\phi_{\rm LUMO}^{\rm mol}|\Sigma|\phi_{\rm LUMO}^{\rm mol}} = 1.023$, $\braket{\phi_{\rm CBM+6}^{\rm TMD}|\Sigma|\phi_{\rm CBM+6}^{\rm TMD}} = 0.445$, $\braket{\phi_{\rm CBM+3}^{\rm TMD}|\Sigma|\phi_{\rm CBM+3}^{\rm TMD}} = 0.446$, and the off-diagonal elements are all close to zero.

Using these values from projection $GW$ calculations of the TMD and of the molecule, one can derive from Eq. \eqref{eq:exp2} that $\braket{\phi_{\rm CBM+2}^{\rm tot}|\Sigma|\phi_{\rm CBM+2}^{\rm tot}} = 0.654$ eV. This is in very good agreement with the interface $GW$ calculation of the CBM+2 of the H$_2$Pc:MoS$_2$ system (the resonance of H$_2$Pc LUMO), whose self-energy is 0.667 eV. 

\bibliography{tmd}

\end{document}